\documentclass[10pt,preprint]{aastex}
\newcommand{\cpd}{CPD~$-$59$\arcdeg$2661}
\begin{document}

\title{OPENING THE TREASURE CHEST: A NEWBORN STAR CLUSTER EMERGES FROM
ITS DUST PILLAR IN CARINA}

\author{Nathan Smith\altaffilmark{1,2,3}}

\affil{Center for Astrophysics and Space Astronomy, University of
Colorado, 389 UCB, Boulder, CO 80309}

\author{Keivan G.\ Stassun}

\affil{Physics \& Astronomy Department, Vanderbilt University, 1807
Station B, Nashville, TN 37235}

\author{John Bally\altaffilmark{2}}

\affil{Center for Astrophysics and Space Astronomy, University of
Colorado, 389 UCB, Boulder, CO 80309}


\altaffiltext{1}{Hubble Fellow; nathans@casa.colorado.edu}

\altaffiltext{2}{Visiting Astronomer, Cerro Tololo Inter-American
Observatory, National Optical Astronomy Observatory, operated by the
Association of Universities for Research in Astronomy, Inc., under
cooperative agreement with the National Science Foundation.}

\altaffiltext{3}{Visiting Astronomer at the New Technology Telescope
of the European Southern Observatory, La Silla, Chile.}

\begin{abstract}

We present detailed observations of the Treasure Chest, a compact
nebula at the head of a dust pillar in the southern Carina nebula.
This object is of interest because it is an example of a dense young
cluster containing at least one massive star, the formation of which
may have been triggered by feedback from the very massive stars in the
Carina nebula, and possibly $\eta$ Carinae itself.  Our observations
include narrowband images of H$\alpha$, [S~{\sc ii}], [O~{\sc iii}],
Pa $\beta$, [Fe~{\sc ii}], and H$_2$, broadband {\it JHK} images, and
a visual-wavelength spectrum of the nebula.  We use these data to
investigate both the nebular and stellar content of the object.  The
near-infrared emission-line images reveal a cavity inside the head of
the dust pillar, which contains a dense cluster of young stars, while
the observed spectrum of the nebula is consistent with an H~{\sc ii}
region ionized by the O9.5 V star \cpd.  The embedded infrared cluster
was discovered in 2MASS data, but our new {\it JHK} images have
improved spatial resolution and sensitivity, allowing an analysis of
the stellar content of the newborn cluster.  After subtracting
contamination of field stars within the Carina nebula itself, we
compare the cluster's color magnitude diagram with pre-main-sequence
isochrones to derive a likely cluster age less than about 0.1 Myr.
This is in reasonable agreement with the dynamical age of a few times
10$^4$ yr for the expanding nebular cavity, indicating extreme youth.
Stars in the Treasure Chest cluster are highly reddened, with
extinction values as large as $A_V\sim$50.  Two-thirds of cluster
members show strong infrared excess colors indicative of circumstellar
disks, which may prove to be among the highest fraction yet seen for a
young cluster once L-band photometry is considered.  All evidence
suggests that the Treaure Chest is an extremely young cluster that is
just now breaking out of its natal cloud into the surrounding massive
star forming region, and is a good target for more detailed study.

\end{abstract}

\keywords{H~{\sc ii} regions --- ISM: individual (NGC~3372) --- stars:
formation --- stars: luminosity function, mass function}

\section{INTRODUCTION}

The Carina nebula (NGC~3372) is a giant H~{\sc ii} region rich in
complex structure that has just recently been recognized as an
important region of ongoing active star formation.  Early surveys of
the central part of the nebula in molecular lines and far-infrared
(IR) continuum suggested that neutral gas and dust is mostly evacuated
from the core of the nebula and that it lacks significant current star
formation (Harvey et al.\ 1979; Ghosh et al.\ 1988; de Grauww et al.\
1981).  It was assumed that radiation and stellar winds from hot
massive stars in Carina are just clearing away the last vestiges of
their natal molecular cloud.

However, recent studies at IR wavelengths have dramatically altered
this view.  Megeath et al.\ (1996) found an embedded near-IR source at
the edge of a dark cloud near $\eta$ Car.  A large-scale thermal-IR
survey with the {\it Midcourse Space Experiment} ({\it MSX}) revealed
dozens of compact IR sources that were suggested as potential sites of
ongoing and possibly triggered star formation (Smith et al.\ 2000).
Many of these are found in the southern part of the nebula at the
heads of giant dust pillars pointing back toward the massive stars in
the core of the nebula (Smith et al.\ 2000; Rathborne et al.\ 2004).
One southern pillar contains a luminous Class~I protostar that drives
the parsec-scale HH666 outflow (Smith et al.\ 2004a), and several
other dust pillars may contain embedded IR sources and even star
clusters seen in 2MASS data (Rathborne et al.\ 2004).  Thus, the dust
pillars in the Carina nebula are akin to the famous pillars in M16,
which contain embedded near-IR sources and may be sites of triggered
star formation (Sugitani et al.\ 2002; Thompson et al.\ 2002;
McCaughrean \& Andersen 2002).  With this ongoing star formation in
the vicinity of some of the hottest and most massive stars known in
the Galaxy (Walborn 1995; Walborn et al.\ 2002), the Carina nebula may
also be an analog of two-stage starburst regions like 30~Dor or
NGC~604.  These types of regions are useful for studying a second
generation of stars, whose formation may have been triggered by
feedback from the first generation.  At a distance of $\sim$2.3 kpc
(Walborn 1995), the Carina nebula provides a laboratory to study this
phenomenon in exquisite detail compared to extragalactic examples.

An object we refer to as the ``Treasure Chest'' (a dust pillar
associated with the star \cpd) is located in the southern part of the
Carina nebula (see Figure 1).  The H$\alpha$ nebulosity surrounding
\cpd\ was first noted by Thackeray (1950), and its visual-wavelength
emission was studied in greater detail by Walsh (1984; see also Herbst
1975; van den Bergh \& Herbst 1975).  Walsh (1984) noted that the
spectral type of \cpd\ was O9.5 V, and that the compact nebula
appeared to be expanding at $\sim$25 km s$^{-1}$.  The nebula and
\cpd\ are coincident with G287.84-0.82, one of the brightest of the
`South Pillars' identified from thermal-IR emission by Smith et al.\
(2000).  Rathborne et al.\ (2004) noted that G287.84-0.82 appeared to
harbor a probable young star cluster seen in 2MASS data (first noted
by Dutra \& Bica 2001), and showed that the IR spectral energy
distribution implied a total integrated luminosity of $\sim$10$^{6.6}$
L$_{\odot}$.  The spectral type of \cpd, the embedded cluster, and the
high IR luminosity imply that the Treasure Chest is a site of recent
{\it massive} star formation. In this paper we take a closer look at
the emission from the dust pillar around \cpd, the young luminous star
cluster embedded within it, and the relationship between them.

\section{OBSERVATIONS AND DATA REDUCTION}

\subsection{Narrow-band Optical Images}

We obtained narrowband images of the southern Carina nebula on 2001
December 18 using the 8192$\times$8192 pixel imager MOSAIC2 mounted at
the prime focus of the Cerro Tololo Interamerican Observatory (CTIO)
4m Blanco telescope.  This camera has a pixel scale of
$\sim$0$\farcs$27 and provides a 35$\farcm$4 field of view, only a
small portion of which is discussed here.  The seeing during the
observations was about 0$\farcs$8.  We used narrowband interference
filters ($\Delta\lambda \; \approx \; 80$ \AA) centered on [O~{\sc
iii}] $\lambda$5007, H$\alpha$ (also transmitting [N~{\sc ii}]
$\lambda$6548 and $\lambda$6583), and [S~{\sc ii}]
$\lambda\lambda$6717,6731.  In each filter, we took several individual
exposures with slight positional offsets to correct for gaps in the
CCD array, and to correct for detector artifacts.  Total exposure
times and other details are listed in Table 1.  We reduced the data in
the standard fashion with the {\sc mscred} package in
IRAF,\footnote{IRAF is distributed by the National Optical Astronomy
Observatories, which are operated by the Association of Universities
for Research in Astronomy, Inc., under cooperative agreement with the
National Science Foundation.} and absolute sky coordinates were
computed with reference to US Naval Observatory catalog stars.
Emission-line images were flux calibrated with reference to {\it
Hubble Space Telescope} images of the Keyhole
Nebula\footnotemark\footnotetext{See {\url
http://oposite.stsci.edu/pubinfo/pr/2000/06/} and Smith et al.\
(2004b).}, obtained from the {\it HST} archive (the position observed
by {\it HST} was included in the MOSAIC2 field of view).  Appropriate
corrections were made for the H$\alpha$ filter, which is wider than
the {\it HST}/WFPC2 F656N filter that mostly excludes the [N~{\sc ii}]
lines, based on line intensities in the H~{\sc ii} region near the
Keyhole (Smith et al.\ 2004b).  Figure 2 shows a 3-color composite
image of \cpd\ and its surroundings made from optical CTIO/MOSAIC2
data.

\subsection{Near-Infrared Images}

Infrared images of the dust pillar and cluster surrounding \cpd\ were
obtained on 2003 March 11 using SOFI, the facility near-IR imager and
spectrograph mounted on the New Technology Telesope (NTT) of the
European Southern Observatory (ESO) at La Silla, Chile.  SOFI uses a
1024$\times$1024 pixel Hawaii HgCdTe array, with a pixel scale of
0$\farcs$288 and a field of view of roughly 4$\farcm$9.  Under
photometric skies with 0$\farcs$7 seeing, we obtained images of the
Treasure Chest in the $J$, $H$, and $K_S$ broadband filters, as well
as narrowband filters isolating Pa$\beta$ at 1.28 $\micron$, [Fe~{\sc
ii}] at 1.64 $\micron$, and H$_2$ 1-0 S(1) at 2.12 $\micron$.  In each
filter, multiple frames were taken at many different positions
(dithering).  Total exposure times and other details are listed in
Table 1.  A sky position roughly 2$\arcdeg$ south of the nebula was
also observed to characterize the sky emission.  The observations were
reduced using standard IR data reduction procedures in {\tt IRAF}.  To
flux calibrate the $J$, $H$, and $K_S$ images, we chose several field
stars included in our images that were near \cpd\ and reasonably well
isolated, and we adopted their fluxes listed in the 2MASS point source
catalog.\footnotemark\footnotetext{{\url
http://irsa.ipac.caltech.edu/2mass.html}} To flux calibrate the
narrowband images, we used the same field stars and interpolated their
continuum flux to the filter wavelength; calibration of the narrow
filters has a roughly 20\% uncertainty in absolute flux.  Figure 3
shows a 3-color composite image of \cpd\ and its surroundings in
near-IR emission lines made from NTT/SOFI data, and Figure 4 shows the
same for the $J$, $H$, and $K_S$ broadband filters.

\subsection{Optical Spectroscopy}

Low-resolution ($R \ \sim \ 700-1600$) spectra from 3600 to 9700 \AA \
were obtained on 2002 March 1 and 2 using the RC Spectrograph on the
CTIO 1.5-m telescope.  Long-slit spectra were obtained with the
1$\farcs$5-wide slit aperture oriented at P.A.$\approx$60$\arcdeg$
offset about 1$\arcsec$ south of \cpd\ as shown in Figure 5.  The pixel
scale in the spatial direction was 1$\farcs$3.  Spectra were obtained
on two separate nights in two different wavelength ranges (blue,
3600-7100 \AA, and red, 6250-9700 \AA), with total exposure times and
other details listed in Table 1.  Sky subtraction was accomplished by
observing a blank sky position roughly 2$\fdg$5 south.  Sky conditions
were mostly photometric, although a few thin transient clouds were
present on the night of March 2 when the blue spectrum was
obtained. Flux calibration and telluric absorption correction were
accomplished using similar observations of the standard stars LTT-3218
and LTT-2415.

Although the long-slit spectra were sky subtracted, the spectrum of
our target is contaminated by bright emission from the background
Carina nebula H~{\sc ii} region.  This background was subtracted using
a fit to several positions in the background sky on either side of the
target (the several background positions are labeled ``B'' in Figure
5).  From the resulting background-subtracted long-slit spectra, we
made an extracted one-dimensional (1-D) spectrum from a segment of the
slit that sampled emission from the compact H~{\sc ii} region
surrounding \cpd\ (see Figure 5).  The blue and red wavelength ranges
of these extracted 1-D spectra were merged to form a single 3600-9700
\AA \ spectrum, with a common dispersion of 2 \AA \ pixel$^{-1}$; the
average of the two was taken in the region of the spectrum near
H$\alpha$ where the blue and red spectra overlapped. A small
correction of about +5\% was made to the absolute flux of the blue
spectrum so that it matched the red in the overlapping region; this
difference was probably due to the very thin transient clouds present
on the second night when the blue spectrum was obtained.  The final
flux-calibrated spectrum is shown in Figure 6.  Uncertainty in the
absolute flux calibration is roughly $\pm$10\%, but our analysis of
the spectrum below relies on relative line fluxes, where uncertainty
in the reddening and measurement errors (for faint lines) dominate the
results.

Observed intensities of many relevant lines are listed in Table 2,
relative to H$\beta$=100.  Uncertainties in these line intensities
vary depending on the strength of the line and the measurement method.
The integrated fluxes of isolated emission lines were measured; for
these, brighter lines with observed intensities greater than 10 in
Table 2 typically have measurement errors of a few percent, and weaker
lines may have uncertainties of $\pm$10 to 15\%.  The uncertainties
increase somewhat at the blue edge of the spectrum.  Blended pairs or
groups of lines were measured by fitting Gaussian profiles.  For
brighter blended lines like H$\alpha$+[N~{\sc ii}] and [S~{\sc ii}],
the measurement uncertainty is typically 5 to 10\%.  Obviously, errors
will be on the high end for faint lines adjacent to bright lines, and
errors will be on the low end for the brightest lines in a pair or
group, or lines in a pair with comparable intensity.  Table 2 also
lists dereddened line intensities.  The reddening used to correct the
observed line intensities was determined by comparing observed
strengths of Hydrogen lines to the Case B values calculated by Hummer
\& Storey (1987).  As shown in Figure 7, the observed Balmer and
Paschen decrements suggest a value for $E(B-V)$ of roughly
0.65$\pm$0.04, using the reddening law of Cardelli, Clayton, \& Mathis
(1989) with $R_V = A_V \div E(B-V) \ \approx \ 4.8$, which is
appropriate for local extinction from dust clouds around the Keyhole
nebula (Smith 1987; Smith 2002).  This is close to the value of
$E(B-V)$=0.6 derived by Walsh (1984) for the star \cpd.  Table 3 lists
representative physical quantities like electron density and
temperature derived from a standard deductive nebular analysis of the
usual line ratios (e.g., Osterbrock 1989).  These are useful to guide
photoionization models described below in \S 4.  The ``model'' line
intensities listed in the last column of Table 2 will be discussed in
\S 4.

\section{IMAGES OF THE NEBULOSITY}

At visual wavelengths (Figure 2), \cpd\ is found roughly at the center
of diffuse emission-line nebulosity extending to radii of
15-20$\arcsec$ (roughly 0.2 pc), as noted by Thackeray (1950) and
Walsh (1984).  Filamentary structure is also seen farther from the
star, especially in [S~{\sc ii}], which appears to outline part of a
dark dust pillar seen as a silhouette in [O~{\sc iii}] and H$\alpha$
in Figure 2.

Images at near-IR wavelengths that penetrate the dust screen clearly
suggest a more interesting morphology.  Instead of a single star
surrounded by diffuse nebulosity, Figures 3 and 4 show that \cpd\ is a
member of a rich cluster of stars in a cavity embedded inside the head
of an externally-illuminated dust pillar.  The wall of the cavity
(seen best in H$_2$ 2.122 $\micron$ emission in Fig.\ 3) has a radius
of roughly 25$\arcsec$ (0.3 pc), which is larger than the nebulosity
seen at visual wavelengths.  We refer to this cluster and the
surrounding nebulosity as the ``Treasure Chest'', because the
morphology is reminiscent of an opened container with sparkling riches
inside.\footnotemark\footnotetext{One usually expects to find a
treasure chest at the bottom of the sea in a sunken pirate ship.
Following this vein, we point out a molecular globule roughly
2$\arcmin$ to the west/southwest of the Treasure Chest, with a
morphology that begs to be called the ``Sea Horse'' nebula.
Additionally, many tadpole-shaped globules are seen throughout the
Carina nebula (Smith et al.\ 2003).}  The embedded cluster and cavity
inside the head of a dust pillar are reminiscent of the ``Mount St.\
Helens'' pillar in 30 Doradus (Walborn 2001), but perhaps at a
somewhat earlier phase just before it has removed its summit.  The
presence of a compact cluster implies ongoing {\it massive} star
formation (consistent with the O9.5 V spectral type of \cpd; Walsh
1984), as compared to the previously documented cases of intermediate-
and low-mass star formation in Carina (Smith et al.\ 2004a, Megeath et
al.\ 1996).

The multiwavelength emission-line structure of the outer edges of the
dust pillar is consistent with an externally-ionized photoevaporative
flow from the surface of a dense molecular cloud (see the discussion
of the Finger globule in the northern part of the Carina nebula; Smith
et al.\ 2004b).  Limb-brightened [S~{\sc ii}] emission is seen at the
edge of the cloud (the ionization front) with more extended H$\alpha$
and [O~{\sc iii}] at larger distances in the ionized evaporative flow.
A thin layer of H$_2$ emission is seen behind the ionization front,
indicating a dusty and geometrically thin photodissociation region
reaching column densities of $n_H\gg$10$^{22}$ cm$^{-2}$.  The eastern
side of the dust pillar has a remarkably straight edge pointing back
toward $\eta$ Carinae and the Tr16 cluster in the heart of the Carina
nebula (see Fig.\ 1), much like the globule associated with HH~666
(Smith et al.\ 2004a), suggesting that $\eta$ Car or other massive
stars in Tr16 were responsible for shaping the Treasure Chest.

Figure 2 indicates severely non-uniform extinction across the field of
view; both on large spatial scales associated with the foreground of
the Carina nebula, and on smaller scales associated with compact
clumps within the dust pillar, superposed on the compact H~{\sc ii}
region around \cpd.  Thus, the value of $E(B-V)$=0.65 that we derived
earlier (Table 3 and Fig.\ 7) indicates a representative average,
while the true reddening may vary strongly with position.

The visual-wavelength emission lines apparently trace only a part of
the ionized cavity that is beginning to break out of its surrounding
cocoon.  Walsh (1984) observed line splitting of $\sim$25 km s$^{-1}$
at the position of the nebula, indicating expansion at roughly $\pm$12
km s$^{-1}$ (near the sound speed).  However, emission-line images
show no clear evidence for outflow activity like highly-collimated
bipolar Herbig-Haro jets (Reipurth \& Bally 2001) or wider-angle
bipolar outflows that would likely be seen in shock tracers such as
[S~{\sc ii}], [Fe~{\sc ii}], or H$_2$ emission.  Nothing like the
dramatic outflow from the BN/KL region of Orion (Salas et al.\ 1999;
Schild et al.\ 1997; Genzel \& Stutzki 1989; Shuping et al.\ 2004) is
seen here, even though the inferred luminosity of $\sim$10$^{6.6}$
L$_{\odot}$ (Rathborne et al.\ 2004) is significantly higher than
BN/KL.

The expansion speed of $\pm$12 km s$^{-1}$ observed by Walsh (1984)
implies an age for the cavity (radius of $\sim$25$\arcsec$ or 0.3 pc)
of a few $\times$10$^4$ yr.  This cavity size is also much smaller
than a typical Str\"{o}mgrem sphere for an O9.5 V star and an ambient
density of $\sim$500 cm$^{-3}$ (Table 3), which would be almost 1 pc.
The young dynamic age and the small size of the cavity compared to the
expected Str\"{o}mgren sphere radius suggest that the cluster around
\cpd\ is extrememly young, and that the cavity is just in the initial
stages of expansion.  Alternatively, the H~{\sc ii} region may be
dusty, and grains may absorb a large fraction of the Lyman continuum
luminosity.  In any case, the compact H~{\sc ii} region is probably
caught in the early phases of breaking out of and destroying the head
of the surrounding dust pillar.  Thus, the H~{\sc ii} region is not in
photoionization equilibrium, which may affect the analysis of the
spectrum in the next section.  Judging by the morphology in Figures 3
and 4, the expansion of the cavity seems to be encountering less
resistance toward the southwest, while a significant reservior of
molecular material impedes its expansion toward the northeast (see
also Rathborne et al.\ 2004).  Note that many of the reddened stars
that are presumably cluster members are also located toward the north
and northeast of \cpd\ near the edge of the cavity (Fig.\ 4).

Finally, we note that the bright star located $\sim$30$\arcsec$
northeast of \cpd\ (at $\alpha_{2000}$=10$^{\rm h}$45$^{\rm
m}$57$\fs$3, $\delta_{2000}$=$-$59$\arcdeg$56$\arcmin$43$\arcsec$) is
probably not a member of the embedded cluster.  This star (Hen
3-485=Wra 15-642) is a Be star and is therefore somewhat evolved, and
so it is probably too old to be associated with the extremely young
cluster around \cpd.  Instead, Massey \& Johnson (1993) considered it
to be a member of Tr16.  It also appears somewhat disconnected from
the bright nebulosity around \cpd, and may be in the foreground of the
Treasure Chest, while still within the confines of the larger Carina
nebula.

\section{SPECTROSCOPIC ANALYSIS OF THE H~{\sc ii} REGION}

To understand the dereddened spectrum of the nebula around \cpd\
quantitatively, we employed the spectral synthesis code {\sc cloudy}
(Ferland 1996), using simplified assumptions about the geometry and
other factors gleaned from the analysis above.  We approximated the
nebula as a 0.5 pc sphere filled with a density\footnote{This density
comes from the electron density in Table 3, derived from the relative
intensities of the [S~{\sc ii}] $\lambda\lambda$6717,6731 lines
observed in the nebula.} $n_H$=500 cm$^{-3}$, and with the density
rising slightly to 600 cm$^{-3}$ at the outer edge.  We used {\sc
cloudy}'s standard H~{\sc ii} region abundances and dust content
(similar to the Orion nebula; see Ferland 1996).  A range of
properties for the ionizing source was tested; the best results were
found using a blackbody with T=31,500~K and L=8$\times$10$^{4}$
L$_{\odot}$.  This is typical for a main-sequence O 9.5 V star, in
agreement with the observed spectral type of \cpd\ (Walsh 1984).  To
adequately match the observed spectrum, however, we needed to adjust
this blackbody by extinguishing roughly half the hydrogen and helium
ionizing photons.  This provided much better results than simply using
a cooler blackbody, for example, which has a different spectral energy
distribution in the UV.  The rationale for the depletion of ionizing
photons in the model might be that the nebula around \cpd\ is not in
equilibrium, while {\sc cloudy} is an equilibrium code.  The cavity's
observed size in Figure 3 is smaller than a Str\"{o}mgren sphere for
an O 9.5 V star, and it is observed to be expanding (Walsh et al.\
1984).  In that case, one can interpret the deficit of ionizing
photons as ionizations not balanced by recombinations as the nebula
increases in size (another way to approximate the observed spectrum
was to simply use a larger radius than observed).  With these input
assumptions, the model produced a fair approximation of the observed
line intensities, and ionizing photons were used-up near the model
nebula's outer boundary as the gas became fully molecular, as expected
for the observed cavity.

However, some discrepancies between the simple model and the observed
spectrum were difficult to reconcile without modifying the chemical
abundances.  Important cooling lines of S and O in the visual spectrum
were too weak, and N lines were too strong.  Therefore, we lowered the
abundance of N by 20\%.  This adjustment allowed the [N~{\sc ii}]
lines to be fit satisfactorally, and also increased the strength of
[S~{\sc ii}], [S~{\sc iii}], and [O~{\sc ii}] lines as they took on
the additional burden of cooling the
nebula.\footnotemark\footnotetext{The Ne abundance was also increased
to account for some additional cooling in order to match the observed
electron temperatures in Table 3; however, this {\it ad hoc}
adjustment could be any additional source of cooling and is in no way
a true indicator of an enhanced Ne abundance.}  The final spectrum
with these slightly modified abundances matched the observed spectrum
quite well, and several important lines predicted by the model are
listed in Table 2 for comparison with the dereddened line intensities.
H and He line intensities are reproduced well; in fact, the perfect
agreement of H$\alpha$ suggests that our reddening value of
$E(B-V)$=0.65 is correct.  Important forbidden cooling lines and
temperature/density diagnostics are also well matched, with the
exceptions noted below.

There are still two unsatisfying discrepancies between our model
predictions in Table 2 and the dereddened spectrum of the nebula
around \cpd.  First, the model underpredicted the strength of [O~{\sc
ii}] $\lambda\lambda$3726,3729 by 14\%.  Other lines in the spectrum
were over- or under-predicted by similar amounts in our model, but
[O~{\sc ii}] $\lambda\lambda$3726,3729 is one of the strongest lines
in the nebula and an important coolant, so the poor agreement is
bothersome.  Simply increasing the oxygen abundance would not fix this
problem, because the increased cooling would dramtically affect the
rest of the spectrum, and it would cause additional problems because
the intensities of the red [O~{\sc ii}] lines are matched quite well
by the current model.  One possible explanation is that our observed
spectrum has relatively large calibration uncertainty at the blue end
of the spectrum, which is enough to account for [O~{\sc ii}]
$\lambda\lambda$3726,3729.  A second (and even more severe)
discrepancy between our model and the observed/dereddened spectrum of
\cpd\ is that our model underpredicts the [O~{\sc iii}] lines by about
a factor of two.  We could not fix this discrepancy in a satisfactory
way with adjustments to the model; increasing the effective
temperature of the ionizing source enough to match the [O~{\sc iii}]
intensities, for example, would dramatically increase the strengths of
He~{\sc i} and [S~{\sc iii}] lines as well (some of which are already
too strong).  The only potential solution to this underestimate of the
model [O~{\sc iii}] lines is that the observed spectrum of the nebula
around \cpd\ may be contaminated by emission from gas along the line
of sight in the Carina nebula outside the dust pillar.  Although we
tried to carefully subtract the background nebular emission, the local
extinction is patchy (Fig.\ 2).  The background Carina nebula has
extremely strong [O~{\sc iii}] lines (e.g., Smith \& Morse 2004; Smith
et al.\ 2004b), and even a small amount of contamination might account
for the discrepancy.

In summary, with the caveats that 1) we needed to make a minor
adjustment to the N abundance, 2) we needed to extinguish some of the
ionizing photons to account for the non-equilibrium state of the
nebula, and 3) we needed to invoke some contamination of the [O~{\sc
iii}] lines from the surrounding Carina nebula, we were able to match
the observed spectrum of the nebula around \cpd\ in a satisfactory way
with a simple geometry that was consistent with the observed
morphology in images, and using an ionizing source consistent with an
O 9.5 V star, which is the observed spectral type of \cpd\ itself
(Walsh 1984).  This proves that \cpd\ is indeed the dominant ionizing
source of the Treasure Chest and is likely a member of the embedded
star cluster seen in IR continuum images, as discussed below in \S 5.

\section{STELLAR CONTENT\label{stellar}}

To conduct a preliminary stellar census of the cluster associated with
the Treasure Chest, we present an analysis of the $JHK$ photometry of
point sources in our broad-band near-IR images (\S 2.2).  We begin by
describing the identification of all stellar sources, including
completeness limits, and our measurement of their $JHK$ magnitudes and
colors. We then place the stars on an $H-K$ vs.\ $K$ color-magnitude
diagram (CMD) and compare their placement with appropriate
pre--main-sequence (PMS) isochrones to derive a likely cluster age.
Next, we study $J-H$ and $H-K$ colors to determine the fraction of
stellar sources with IR excesses indicative of circumstellar
disks. Finally, we present a K-band luminosity function (KLF) for the
cluster which will serve as the basis for a follow-up study of the
cluster's initial mass function.

\subsection{$JHK$ photometry\label{photometry}}

We performed standard point-spread-function (PSF) photometry on our
$JHK_S$ images using the IRAF {\sc daophot} package.  Stellar point
sources were identified with {\sc daofind} using a signal-to-noise
ratio of 5. An empirical, spatially variable, model PSF was
constructed separately for each image using $\approx 20$ bright and
relatively isolated stars. To construct the cleanest model PSF
possible, we disregarded stars at the very center of the images, where
the nebular emission is strongest. At the same time, it was necessary
to exclude PSF stars near the edges of the field of view because the
PSF is highly spatially variable and it is in the center of the images
where the PSF is best and where the stars of greatest interest
reside. Thus, we carefully selected PSF stars just at the periphery of
the strong nebular background.

Owing to the strong and spatially variable nature of the nebular
background emission, we performed the PSF photometry in an iterative
fashion with the aim of accurately modeling and subtracting this
background. First, we subtracted all stellar sources identified by
{\sc daofind}, and then interpolated the background over the positions
of the subtracted sources, to produce a model of the nebular
background alone. This background was then subtracted from the
original image, and the PSF photometry performed anew on this
background-subtracted image, with the background in the PSF fitting
now fixed at zero.  We were not able to derive photometry for \cpd\
itself, since this bright star was saturated in our images.

Stellar fluxes derived from the PSF fitting were converted to $JHK$
magnitudes on the CIT system using the photometric zero-points and
color terms determined by the 2MASS\footnote{See
\url{http://www.ipac.caltech.edu/2mass/releases/allsky/doc/sec6\_4b.html}.}
project. From inspection of histograms of the $JHK$ magnitudes so
derived, we find that we are sensitive to $JHK$ magnitudes of 20.4,
19.2, and 18.1, respectively.  To determine our completeness limits in
$JHK$, we used the IRAF package {\sc artdata} to add artificial stars
of varying brightness to the original images. We successfully
recovered 90\% of these artificial stars at $JHK$ magnitudes of 19.3,
18.2, and 17.2, respectively. These then define our 90\% completeness
limits.

\subsection{Cluster boundary and ``background" region\label{boundary}}

In order to distinguish the properties of the young stars in the
Treasure Chest from those of the young stellar population likely
associated with the surrounding star-forming region, we defined a
cluster boundary by tracing the edge of the nebular emission visible
in our narrow-band images (\S 3; Fig.\ 3).  This marks the boundary of
the surrounding elephant trunk that harbors the embedded cluster.

We also defined a circular annulus around this cluster boundary as a
``background" region for comparison. To be sure, this region does not
sample a truly background stellar population, in the sense that this
region is likely to be dominated by young stars in the Carina
star-forming region, not just foreground and background field
stars. But defining the background region in this way will allow us to
characterize the stellar population in the Treasure Chest cluster
separately from that of other young stars in the immediate vicinity of
the cluster.  In the analysis that follows, we will differentiate
between stars in the ``cluster" and ``background" regions so
defined.

\subsection{Stellar number density\label{density}}

From our $K_S$ image we can determine the surface number density of
stellar point sources in the Treasure Chest cluster. At a distance of
2.3 kpc, our image scale of 0$\farcs$27 pixel$^{-1}$ corresponds to
0.003 pc pixel$^{-1}$. We have measured the surface number density by
counting stars in boxes 33$\times$33 pixels, corresponding to
0.1$\times$0.1 pc. In the background annulus region, we detect an
average stellar surface number density of 250 pc$^{-2}$, while in the
cluster region itself we detect a maximum of 1020 pc$^{-2}$ in the
northern part of the cluster.

This is almost certainly a lower limit to the true cluster 
density, as even in our $K_S$ image we can see clear traces of
extinction, and in our analysis below we find stars whose colors
indicate significant extinction, $A_V > 40$ (see \S \ref{cmd}).
Deeper images at 2 $\mu$m will presumably detect even more
sources, although at our current completeness limit of 
$K \approx 17.2$ (\S \ref{photometry}), the KLF for the cluster
region already blends into the KLF of the background region
(see \S \ref{klf}).

Nonetheless, we can say here that the maximum stellar surface number
density in the Treasure Chest is at least $\approx 770$ pc$^{-2}$
(i.e., from 1020--250, as noted above), which is comparable to that
found in other rich young clusters, such as IC348 (Lada \& Lada 1995),
though probably not as high as that found in NGC~2024 (Haisch et al.\
2000) and the Trapezium (Hillenbrand \& Hartmann 1998).

\subsection{Color-magnitude diagram\label{cmd}}

Within the cluster boundary defined above, we detected 172 point
sources in our $J$ image, 194 in our $H$ image, and 199 in our $K_S$
image. Of these, 156 are common to all three images, and 183 are
detected in both the $H$ and $K_S$ images.

The $H-K$ vs.\ $K$ CMD for stars in the cluster and background regions
is shown in Figure 8. Stars located within the cluster region are
displayed as filled points. Dot-dashed lines represent isochrones for
stars with masses from 0.02 M$_\odot$ to 3.0 M$_\odot$, at ages of 0.1
Myr, 1 Myr, and 100 Myr, from the PMS models of D'Antona \& Mazzitelli
(1997), assuming a cluster distance of 2.3 kpc. The isochrones have
been transformed from the $T_{\rm eff}/L$ plane to the $(H-K)/K$ plane
using the relationship between $T_{\rm eff}$ and $HK$ bolometric
corrections compiled by Muench et al.\ (2002). Dashed lines are
reddening vectors for stars with masses of 2.5 M$_\odot$, 0.3
M$_\odot$, 0.08 M$_\odot$, and 0.05 M$_\odot$, and extinctions $A_V$
of 70, 40, 20, and 10 mag, respectively. These extinction vectors
assume the reddening law of Bessell \& Brett (1988) and a ratio of
total-to-selective extinction, $R_V$, of 4.8 (see Smith 2002).  Dotted
lines represent our sensitivity and completeness limits.  Crosses
along the right side of the figure represent typical observational
error bars at various $K$ magnitudes.  Our photometry is evidently
complete for a star at the hydrogen-burning limit (0.08 M$_\odot$), at
an age of 0.1 Myr, seen through $A_V = 20$ mag of extinction.

The CMD of the Treasure Chest cluster exhibits several interesting
features that provide some insight into the nature of its stellar
population. First, the majority of sources in the cluster region
(filled circles) show very red $H-K$ colors, consistent with a young
population of stars still embedded in significant quantities of
intervening dust. Indeed, we find stars with extinctions as large as
$A_V \sim 50$.  This population of highly reddened stars exists in the
background annulus region as well (open circles), indicating the
presence of significant numbers of young stars in the surrounding
star-forming region.

Comparison with the PMS isochrones of D'Antona \& Mazzitelli (1997)
provides further indications of extreme stellar youth.  For stellar
magnitudes down to $K \sim 16$, the cluster stars (filled circles)
display a marked blue ``edge" that is well traced by the 0.1 Myr
isochrone, with a marked decrease in the number of stars blueward of
that isochrone.  For $K > 16$, the 0.1 Myr isochrone begins to merge
in $H-K$ color with the 1 Myr isochrone, which may be more
representative of the young stellar population associated with the
surrounding star-forming region. At these faint magnitudes,
particularly at $K > 16.5$, we also begin to see a significant number
of stars in both the cluster and background regions with blue $H-K$
colors indicative of main-sequence and giant field stars.

\subsection{Color-color diagram\label{ccd}}

In Figure 9 we show a $J-H$ vs.\ $H-K$ color-color diagram for the
stars that we detected in all three passbands. The upper panel shows
stars within the cluster boundary defined above, while the lower panel
shows stars within the background annulus.  Solid lines represent the
colors of main-sequence stars and giants from Bessell \& Brett (1988),
transformed to the CIT system.  Dashed lines represent reddening
vectors emanating from the extrema of the main-sequence and giant
colors. Stars within these reddening vectors are consistent with
reddening due to intervening interstellar dust, although small amounts
of reddening due to circumstellar material cannot be ruled out.

In both panels, red points represent stars redward of the 1 Myr
isochrone in the CMD (Fig.\ 8), whereas green points represent stars
blueward of this isochrone, and which are therefore unlikely to be
young stars associated with the Treasure Chest or the surrounding
star-forming region. Indeed, with few exceptions the green points
follow the main-sequence and giant color relations closely.

The dash-dotted line is the locus of classical T~Tauri stars from
Meyer et al.\ (1997). Stars on or above this line and to the right of
the dashed lines are those with IR excesses indicative of massive
circumstellar disks.
Both the cluster and background regions show evidence for stars with
disks. However, such stars are found in higher proportion in the
cluster region. In the cluster region, 67\% of likely cluster members
(red points) show evidence for circumstellar disks; in the background
region this fraction is 44\%.

The large disk fraction found for the cluster region is almost
certainly a lower limit to the true disk fraction, since $JHK$
photometry is not the most sensitive tracer of disks. For example, in
NGC~2024 (age $<1$ Myr) Haisch et al.\ (2000) found a disk fraction of
$\sim 60\%$ from analysis of its $JHK$ excess fraction, but found a
much higher disk fraction of $\sim 90\%$ when they included $L$-band
photometry in the analysis. That the disk fraction we find in the
Treasure Chest is larger than that found in NGC~2024 via $JHK$
photometry suggests that the true disk fraction in this cluster may
prove to be among the highest yet seen for a young cluster, and
further corroborates the extremely young age from our analysis of the
CMD (\S \ref{cmd}).

\subsection{$K$-band luminosity function\label{klf}}

One of our principal aims in studying the Treasure Chest is ultimately
to measure the mass spectrum of an extremely young cluster that may
have been triggered by an earlier episode of nearby, massive star
formation. One of the primary means for determining the initial mass
function (IMF) of a young cluster is by analysis of its KLF (Muench et
al.\ 2002).

The KLF of the Treasure Chest cluster is shown in Figure 10,
which includes only stars redward of the 0.1 Myr isochrone in the CMD.
The dashed histogram shows the corresponding KLF for stars in the
background annulus region, scaled to the same spatial area as the
cluster region.

The KLF shows a clear excess of stars over the background for all $K$
down to our completeness limit.  We can place a lower limit on the
number of members in the Treasure Chest cluster by summing over the
cluster KLF after subtracting the background KLF.
We find a lower limit of 69 stars down to our completeness limit.
For $K \gtrsim 17$, the cluster KLF merges with the KLF of the
surrounding star-forming region. Thus, while deeper imaging of the
cluster may identify additional stars, it may prove difficult to
statistically separate members of the Treasure Chest cluster from
members of the surrounding nebula for $K \gtrsim 17$ ($M < 0.05$
M$_\odot$ with $A_V = 10$ at an age of $\lesssim 1$ Myr).

\section{SUMMARY AND CONCLUSIONS}

We have undertaken a detailed observational analysis of a dust pillar
in the Carina nebula called the Treasure Chest, as well as its
associated embedded star cluster.  Narrowband images,
visual-wavelength spectra, and broadband near-IR photometry point
toward the following main conclusions, all of which provide
independent evidence of extreme youth:

1.  Emission-line images of the Treasure Chest reveal an embedded star
    cluster occupying a cavity inside the head of an
    externally-ionized dust pillar, and part of the embedded compact
    H~{\sc ii} region appears to be breaking out of the dust pillar
    into the surrounding giant H~{\sc ii} region.  The dust pillar
    points toward $\eta$ Carinae and other stars in the Tr16 cluster,
    raising suspicion that the birth of this young cluster was
    triggered by feedback from nearby massive stars.

2.  The visual-wavelength spectrum of ionized gas from the cavity that
    is breaking through the dust cocoon is ionized by a late O-type
    star, consistent with the spectral type of O~9.5~V derived for the
    central star \cpd\ by Walsh (1984).

3.  Analysis of the color-magnitude diagram of the embedded cluster
    suggests that it has an age $\la$0.1 Myr, which is in reasonable
    agreement with the dynamical age of the expanding nebular cavity
    (a few times 10$^4$ yr).

4.  Stellar photometry also reveals several members of the cluster
    that are highly embedded, with extinction values as high as
    $A_V\sim$50.  This is much higher than the average extinction for
    the ionized gas or \cpd\ itself, suggesting that \cpd\ and its
    compact H~{\sc ii} region are breaking out of the dust pillar on
    the side facing the Earth, while star formation in the cloud
    may be continuing on the far side.

5.  Two-thirds of the cluster members show strong IR excess emission
    indicative of cirumstellar disks, but the fraction may be much
    higher if longer-wavelength data are considered.  Thus, the disk
    fraction for the Treasure Chest cluster may be among the highest
    yet seen for any young clusters.  Some field stars outside the
    cluster (but still within the Carina nebula) also show IR excess
    indicative of young disks, but the fraction is lower than in the
    cluster itself.

We have also provided a preliminary K-band luminosity function, which
suggests that new data with higher spatial resolution and sensitivity
will allow us to accurately measure the cluster's mass function.  The
mass function of this particular cluster will be of great interest,
because it is a ``second-generation'' cluster at the periphery of a
giant H~{\sc ii} region.

\acknowledgements  \scriptsize

Support for N.S.\ was provided by NASA through grant HF-01166.01A from
the Space Telescope Science Institute, which is operated by the
Association of Universities for Research in Astronomy, Inc., under
NASA contract NAS~5-26555.  NOAO funded N.S.'s travel to Chile and
accommodations while at CTIO.  Additional support was provided by NSF
grant AST 98-19820 and NASA grants NCC2-1052 and NAG-12279 to the
University of Colorado.



\begin{deluxetable}{lllcl}
\tabletypesize{\scriptsize}
\tighten
\tablewidth{0pt}
\tablecaption{Observations of the Treasure Chest}
\tablehead{
  \colhead{Telescope} &\colhead{Instrument} &\colhead{Filter or}
       &\colhead{Exp.\ Time} &\colhead{Comment} \\
  \colhead{\ } &\colhead{\ } &\colhead{Emiss.\ Lines} &\colhead{(sec)} 
       &\colhead{\ }       }
\startdata
CTIO 4m   &MOSAIC  &[O~{\sc iii}] $\lambda$5007            &240  &image \\
CTIO 4m   &MOSAIC  &H$\alpha$, [N~{\sc ii}]                &600  &image \\
CTIO 4m   &MOSAIC  &[S~{\sc ii}] $\lambda\lambda$6717,6731 &480  &image \\
NTT       &SOFI    &J                                      &150  &image \\
NTT       &SOFI    &H                                      &150  &image \\
NTT       &SOFI    &K                                      &150  &image \\
NTT       &SOFI    &Pa$\beta$                              &900  &image \\
NTT       &SOFI    &[Fe~{\sc ii}] $\lambda$16435           &900  &image \\
NTT       &SOFI    &H$_2$ 1-0 S(1) $\lambda$21218          &900  &image \\
CTIO 1.5m &RC Spec &blue; 3600-7100 \AA &1200  &long-slit, P.A.=69$\arcdeg$ \\
CTIO 1.5m &RC Spec &red;  6250-9700 \AA &1200  &long-slit, P.A.=69$\arcdeg$ \\
\enddata
\end{deluxetable}

\begin{deluxetable}{llccc}
\tabletypesize{\scriptsize}
\tighten
\tablewidth{0pt}
\tablecaption{Observed, Dereddened, and Model Line Intensities}
\tablehead{
 \colhead{$\lambda$(\AA)} &\colhead{I.D.} &\colhead{(Obs.)} &\colhead{(Dered.)} &\colhead{(Model)}}
\startdata
3726,9	&[O~{\sc ii}]	      &132	      &237	&207	 \\
3798	&H10      	      &2.2	      &3.9	&\nodata \\
3835	&H9      	      &3.0	      &5.1	&\nodata \\
3869	&[Ne~{\sc iii}]	      &2.6	      &4.4	&\nodata \\
3889	&H8, He~{\sc i}	      &8.7	      &14.6	&13.3	 \\
3970	&H$\epsilon$	      &8.0	      &12.8	&\nodata \\
4069,76	&[S~{\sc ii}]	      &2.7	      &4.3	&3.9	 \\
4102	&H$\delta$	      &18.1	      &27.3	&26.3	 \\
4340	&H$\gamma$	      &36.1	      &48.3	&47.0	 \\
4861	&H$\beta$	      &100	      &100	&100	 \\
4959	&[O~{\sc iii}]	      &12.1	      &11.4	&5.5	 \\
5007	&[O~{\sc iii}]	      &35.9	      &33.0	&15.8	 \\
5041	&Si~{\sc ii}	      &1.2	      &1.1	&\nodata \\
5056	&Si~{\sc ii}	      &0.89	      &0.79	&\nodata \\
5199	&[N~{\sc i}]	      &1.4	      &1.1	&\nodata \\
5755	&[N~{\sc ii}]	      &1.4	      &0.89	&0.8     \\
5876	&He~{\sc i}	      &5.0	      &3.0	&3.7     \\
6300	&[O~{\sc i}]	      &2.8	      &1.4	&1.0     \\
6312	&[S~{\sc iii}]	      &0.97	      &0.49	&0.6     \\
6347	&Si~{\sc ii}	      &1.4	      &0.70	&\nodata \\
6364	&[O~{\sc i}]	      &1.6	      &0.80	&\nodata \\
6371	&Si~{\sc ii}	      &0.88	      &0.43	&\nodata \\
6548	&[N~{\sc ii}]	      &87.4	      &40.3	&39.9    \\
6563	&H$\alpha$	      &643	      &295	&295     \\
6583	&[N~{\sc ii}]	      &249	      &113	&117     \\
6678	&He~{\sc i}	      &2.4	      &1.0	&1.0     \\
6717	&[S~{\sc ii}]	      &61.4	      &26.6	&26.8    \\
6731	&[S~{\sc ii}]	      &58.5	      &25.3	&25.4    \\
7065	&He~{\sc i}	      &1.3	      &0.49	&0.7     \\
7136	&[Ar~{\sc iii}]	      &5.6	      &2.1	&3.1     \\
7319,20	&[O~{\sc ii}]	      &5.2	      &1.8	&1.9     \\
7330,31	&[O~{\sc ii}]	      &4.7	      &1.6	&1.6     \\
7612	&[Fe~{\sc ii}]	      &15.1	      &4.7	&\nodata \\
7751	&[Ar~{\sc iii}]	      &1.6	      &0.48	&0.7     \\
8446	&O~{\sc i}	      &9.7	      &2.3	&\nodata \\
8665	&Pa13		      &4.9	      &1.1	&\nodata \\
8750	&Pa12		      &6.9	      &1.5	&\nodata \\
8863	&Pa11		      &7.4	      &1.5	&\nodata \\
9015	&Pa10		      &18.9	      &3.7	&\nodata \\
9069	&[S~{\sc iii}]	      &108	      &20.8	&27.1    \\
9229	&Pa9		      &17.5	      &3.2	&\nodata \\
9532	&[S~{\sc iii}]	      &352	      &59.8	&67.2    \\
9546	&Pa8		      &19.9	      &3.4	&\nodata \\
9615	&Fe~{\sc ii}	      &5.1	      &0.8	&\nodata \\
9711	&[Fe~{\sc ii}]	      &13.3	      &2.2	&\nodata \\
\enddata
\end{deluxetable}

\begin{deluxetable}{llc}
\tighten
\tablewidth{0pt}
\tablecaption{Parameters derived from Dereddened Line Intensities}
\tablehead{
  \colhead{Parameter} &\colhead{units} &\colhead{Value}  }
\startdata
E(B$-$V)                   &mag.       &0.65          \\
n$_{\rm e}$ [S~{\sc ii}]   &cm$^{-3}$  &490$\pm$100   \\
T$_{\rm e}$ [N~{\sc ii}]   &K          &8100$\pm$650  \\
T$_{\rm e}$ [O~{\sc iii}]  &K          &$<$20,000     \\
T$_{\rm e}$ [S~{\sc iii}]  &K          &7000$\pm$700  \\
\enddata
\end{deluxetable}

\begin{figure}
\epsscale{0.9}
\plotone{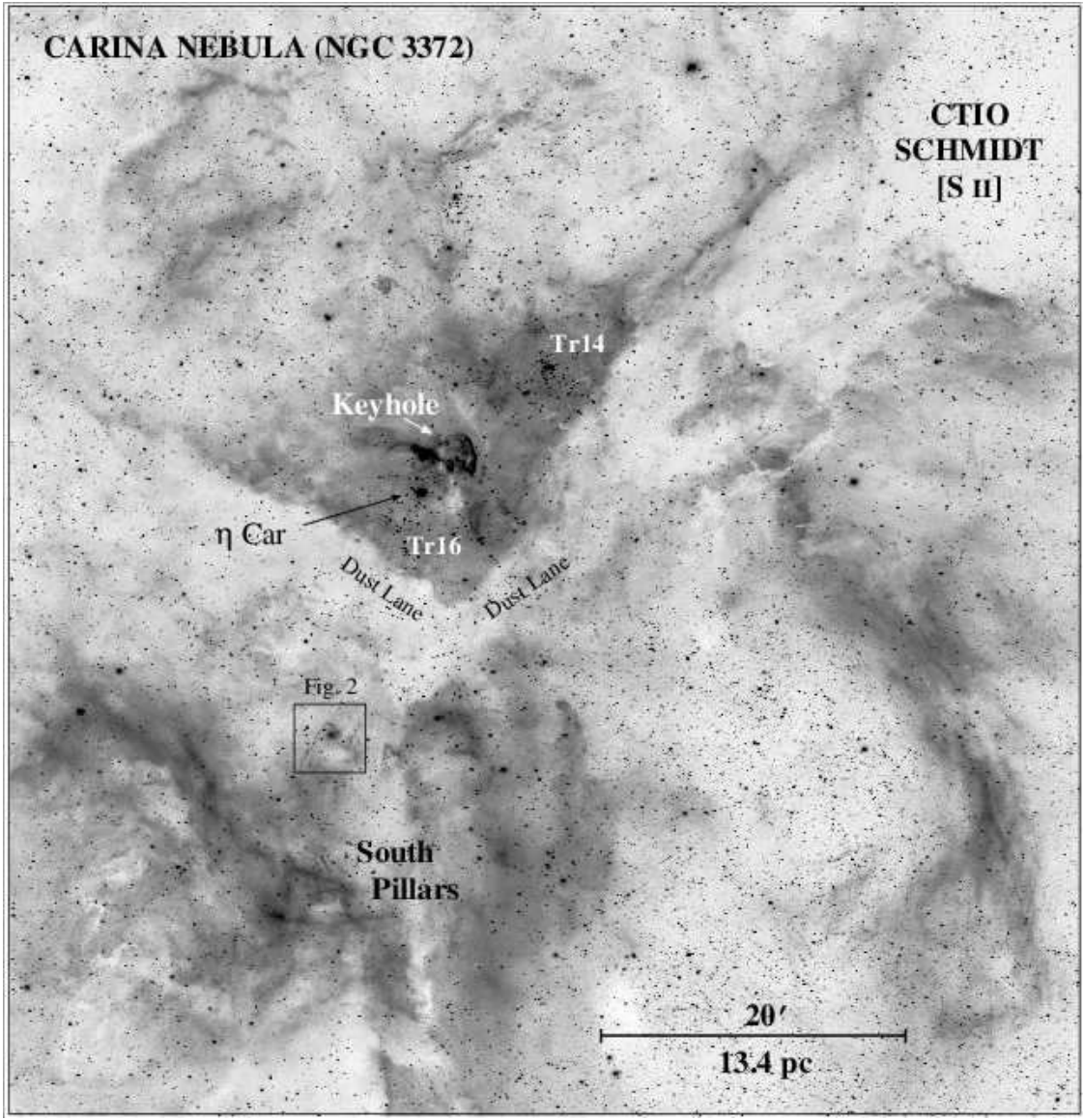}
\caption{ The small box below the center of the image includes \cpd\
itself and denotes the field-of-view of the optical image in Figure 2.
The background image of the large-scale H~{\sc ii} region is in the
light of [S~{\sc ii}] $\lambda\lambda$6717,6731 and was obtained using
the CTIO/Schmidt telescope.  A color version of the image is available
from http://www.noao.edu/image\_gallery/html/im0667.html.  }
\end{figure}

\begin{figure}
\epsscale{0.99}
\plotone{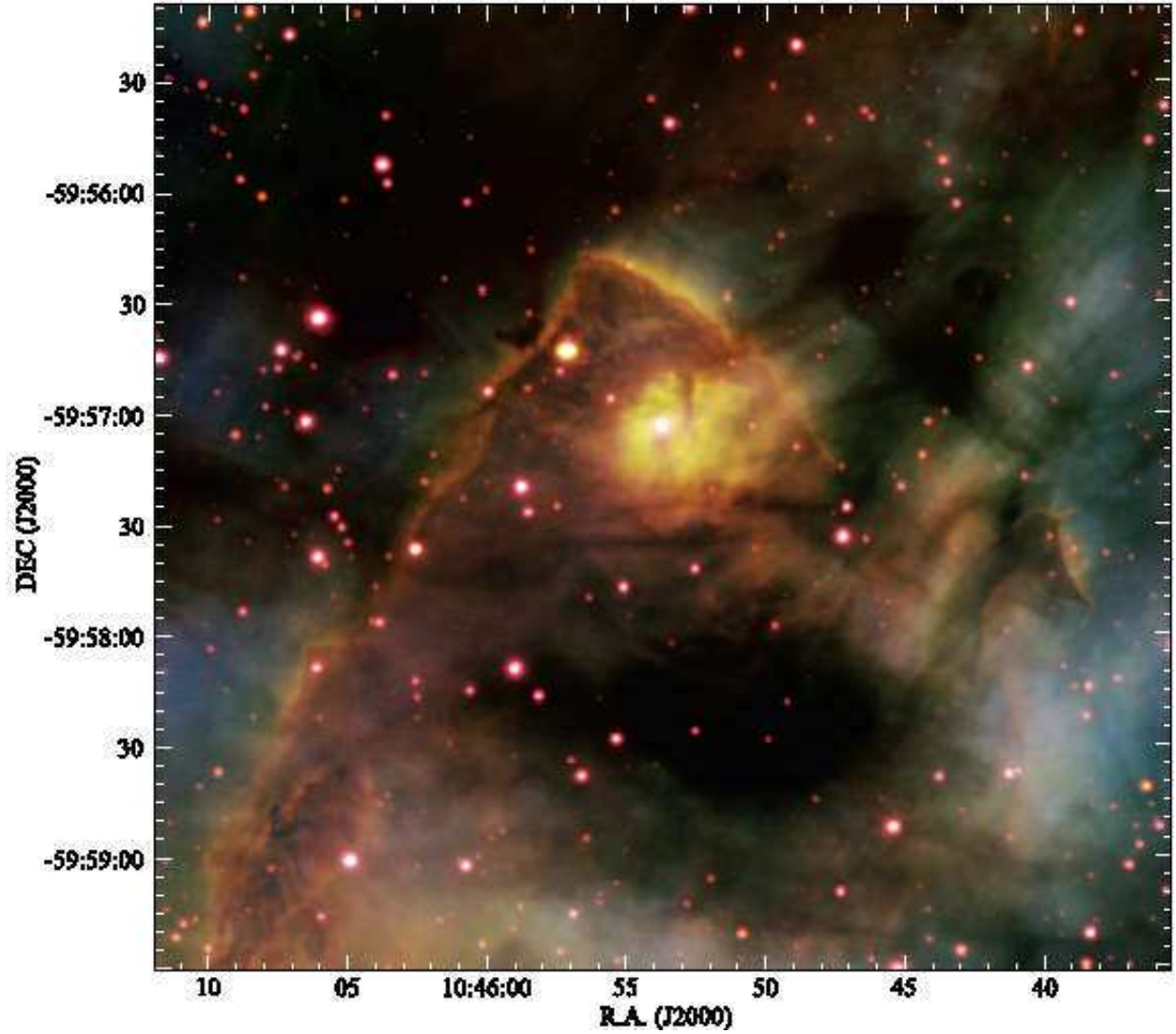}
\caption{Optical image of the environment around \cpd\ taken with the
MOSAIC2 camera on the CTIO 4m telescope, corresponding to the small
box in Figure 1.  Blue is [O~{\sc iii}] $\lambda$5007 emission, green
is H$\alpha$ (and [N~{\sc ii}] $\lambda$6583), and red is [S~{\sc ii}]
$\lambda\lambda$6717,6731. \cpd\ is the star at the center of the small
nebulosity near the center of the image, at $\alpha_{2000}$=10$^{\rm
h}$45$^{\rm m}$53$\fs$7,
$\delta_{2000}$=$-$59$\arcdeg$57$\arcmin$04$\arcsec$.}
\end{figure}

\begin{figure}
\epsscale{0.9}
\plotone{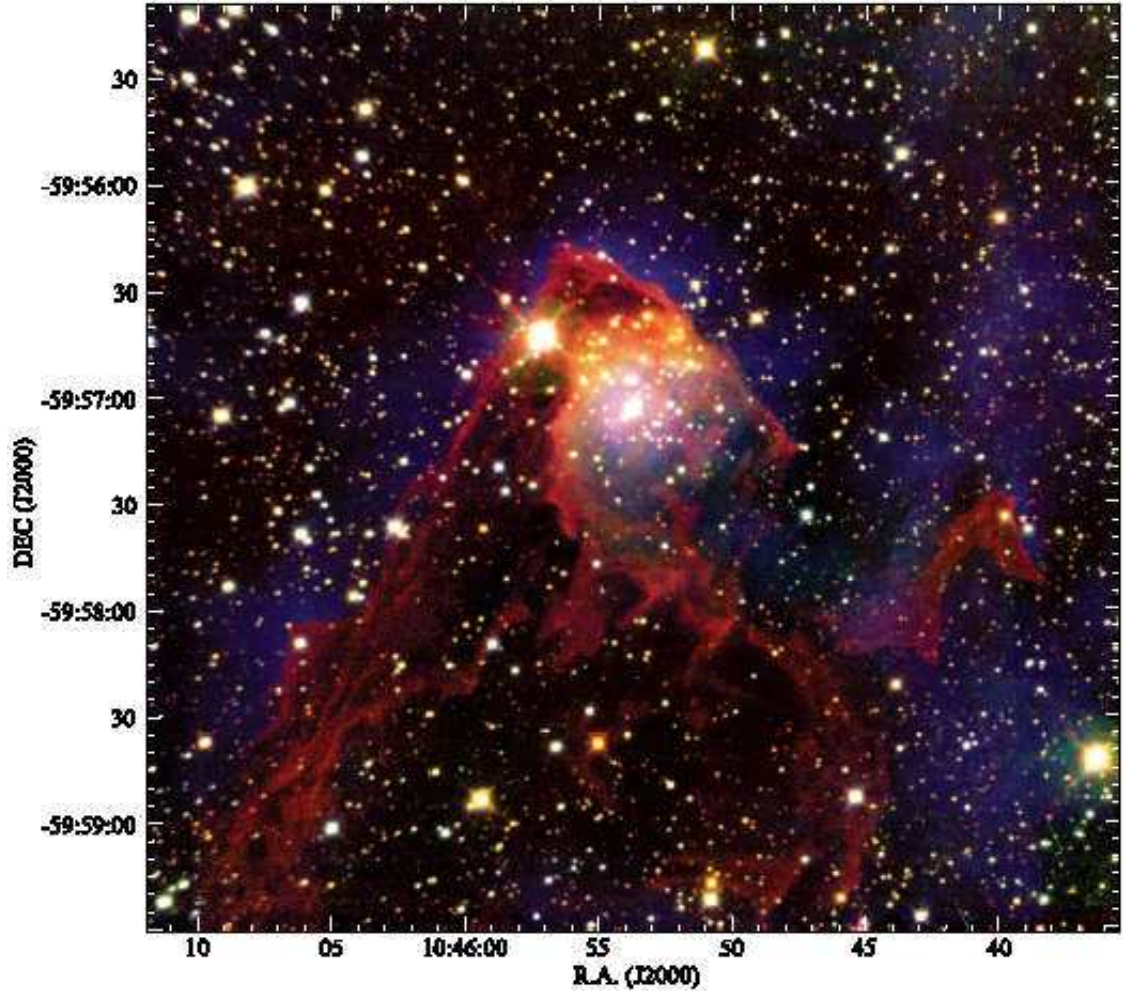}
\caption{Near-IR emission-line image of the environment around \cpd\
taken with SOFI on the NTT.  Blue is Pa$\beta$ $\lambda$12818, green
is [Fe~{\sc ii}] $\lambda$16435, and red is H$_2$ 1$-$0 S(1)
$\lambda$21218.  This image covers the same spatial region as Figure
2.}
\end{figure}

\begin{figure}
\epsscale{0.9}
\plotone{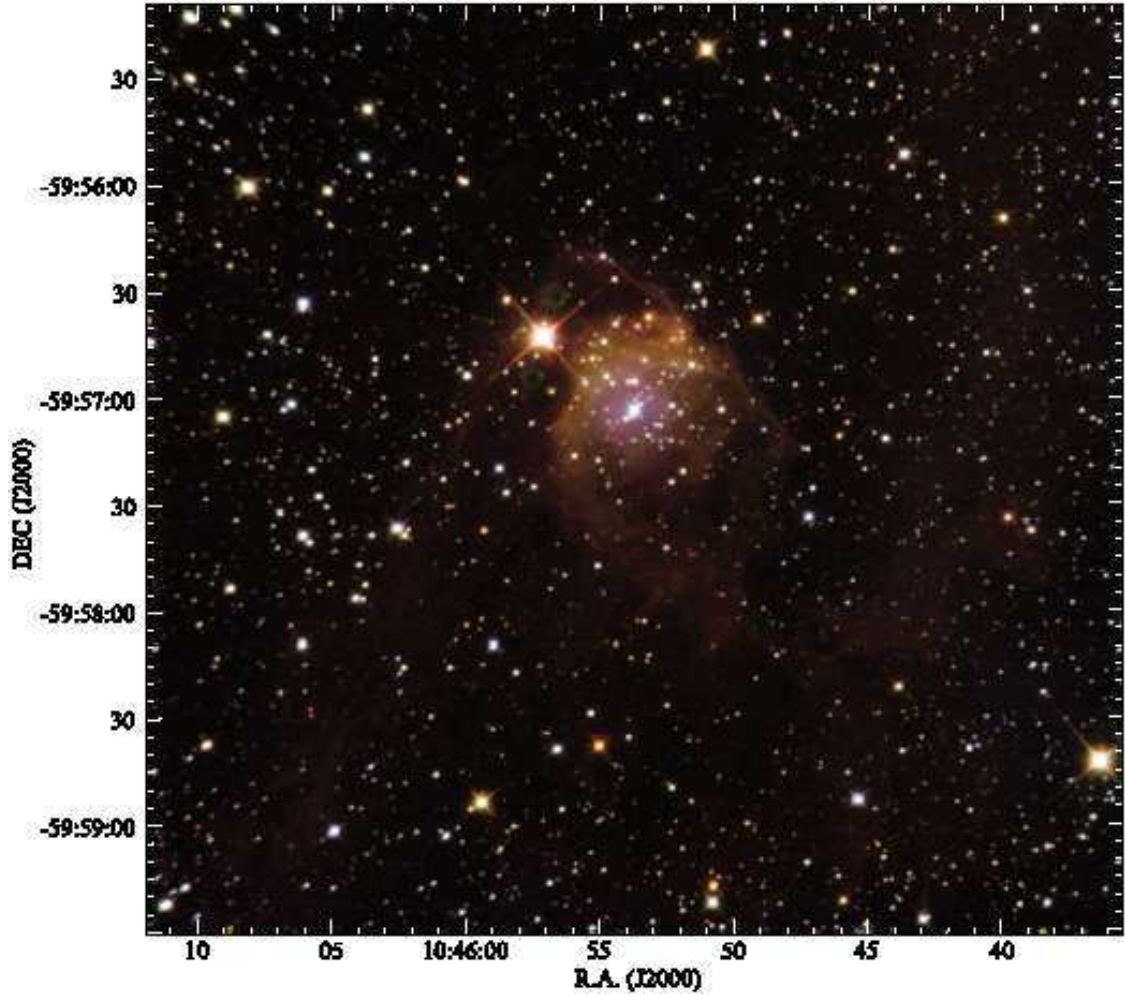}
\caption{Near-IR continuum image of the environment around \cpd\ taken
with SOFI on the NTT.  The J, H, and K broadband filters are shown in
blue, green, and red, respectively.  This image covers the same
spatial region as Figures 2 and 3.}
\end{figure}

\begin{figure}
\epsscale{0.55}
\plotone{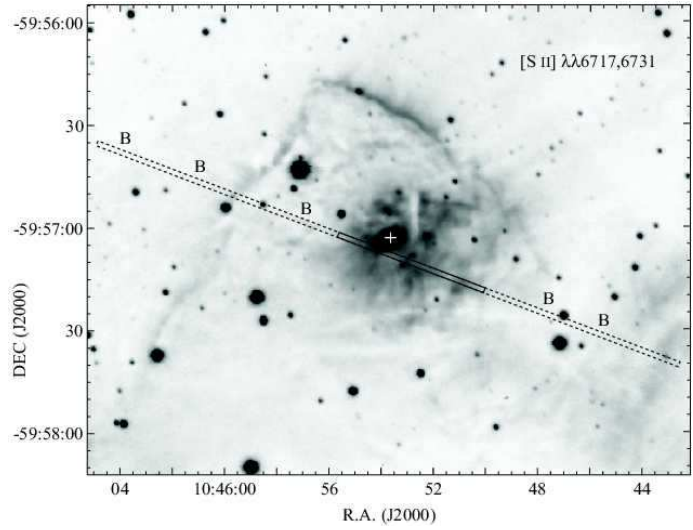}
\caption{The head of the dust pillar surrounding \cpd\ in [S~{\sc
ii}].  The position and orientation of the spectroscopic aperture is
shown with the long dashed box.  The subaperture used to extract the
spectrum in Figure 6 is shown with the solid box, and positions
labeled ``B'' are samples of the background Carina nebula emission
that were fit and subtracted from the spectrum of the compact H~{\sc
ii} region around \cpd, which is marked with a cross.}
\end{figure}

\begin{figure}
\epsscale{0.8}
\plotone{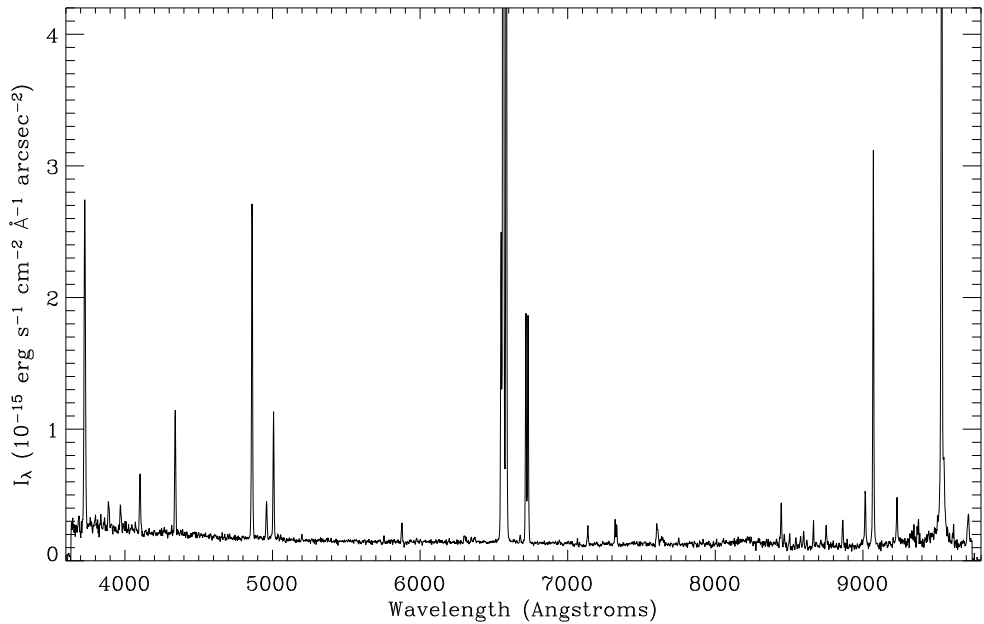}
\caption{The observed optical spectrum of the compact H~{\sc ii}
region around \cpd, not corrected for reddening and extinction.  The
intensity scale shows the average surface brightness integrated over
the section of the slit shown in Figure 5.}
\end{figure}

\begin{figure}
\epsscale{0.5}
\plotone{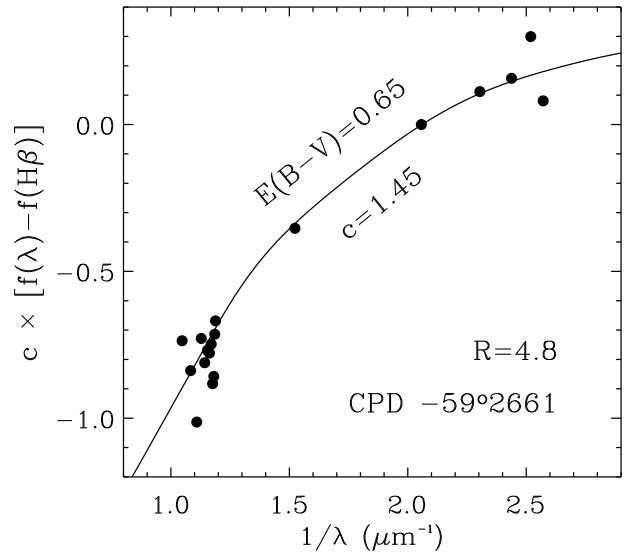}
\caption{Observed reddening for hydrogen lines, relative to H$\beta$
(dots).  The solid curve shows the reddening for $E(B-V)$=0.65 or a
logarithmic extinction at H$\beta$ of c$\approx$1.45 (assuming the
ratio of total to selective extinction is $R_V$=4.8).  Some of the
lines deviate from the curve because they are contaminated by emission
from other lines; H8+He~{\sc i} $\lambda$3889 is a good example.}
\end{figure}

\begin{figure}
\epsscale{0.7}
\plotone{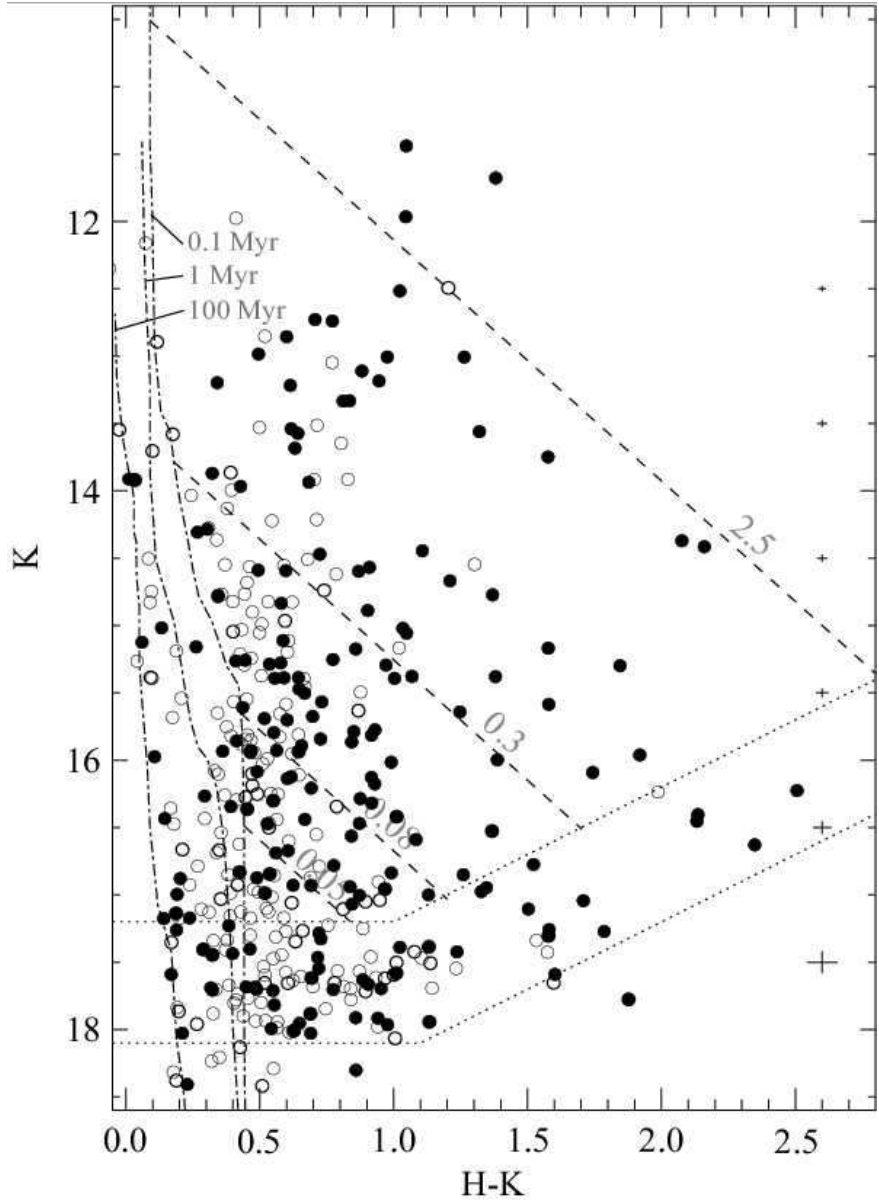}
\caption{Infrared $H-K$ vs.\ $K$ color-magnitude diagram for stars in
the Treasure Chest cluster (filled circles) and background stars in
the surrounding annulus (unfilled circles). Dot-dashed lines represent
isochrones for stars with masses from 0.02 M$_\odot$ to 3.0 M$_\odot$,
at ages of 0.1 Myr, 1 Myr, and 100 Myr, from the PMS models of
D'Antona \& Mazzitelli (1997). Dashed lines are reddening vectors for
stars with masses of 2.5 M$_\odot$, 0.3 M$_\odot$, 0.08 M$_\odot$, and
0.05 M$_\odot$, and extinctions $A_V$ of 70, 40, 20, and 10 mag,
respectively.  Dotted lines represent our sensitivity and completeness
limits.  Crosses along the right side of the figure represent typical
observational error bars at various $K$ magnitudes (see text \S 5.4).}
\end{figure}

\begin{figure}
\epsscale{0.7}
\plotone{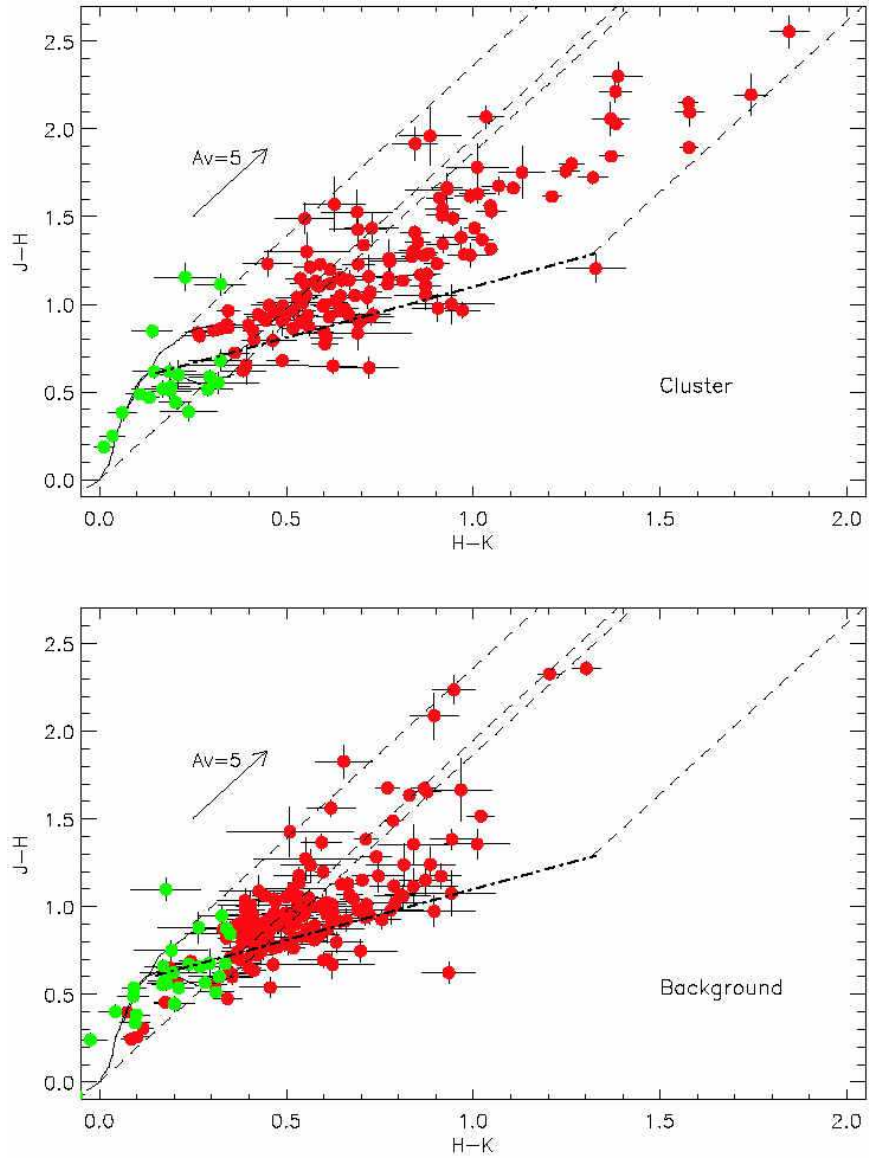}
\caption{The upper panel shows the infrared color-color diagram for
stars within the cluster boundary, while the lower panel shows stars
within the background annulus.  Solid lines represent the colors of
main-sequence stars and giants from Bessell \& Brett (1988),
transformed to the CIT system.  Dashed lines represent reddening
vectors emanating from the extrema of the main-sequence and giant
colors (see text \S 5.5).}
\end{figure}

\begin{figure}
\epsscale{0.8}
\plotone{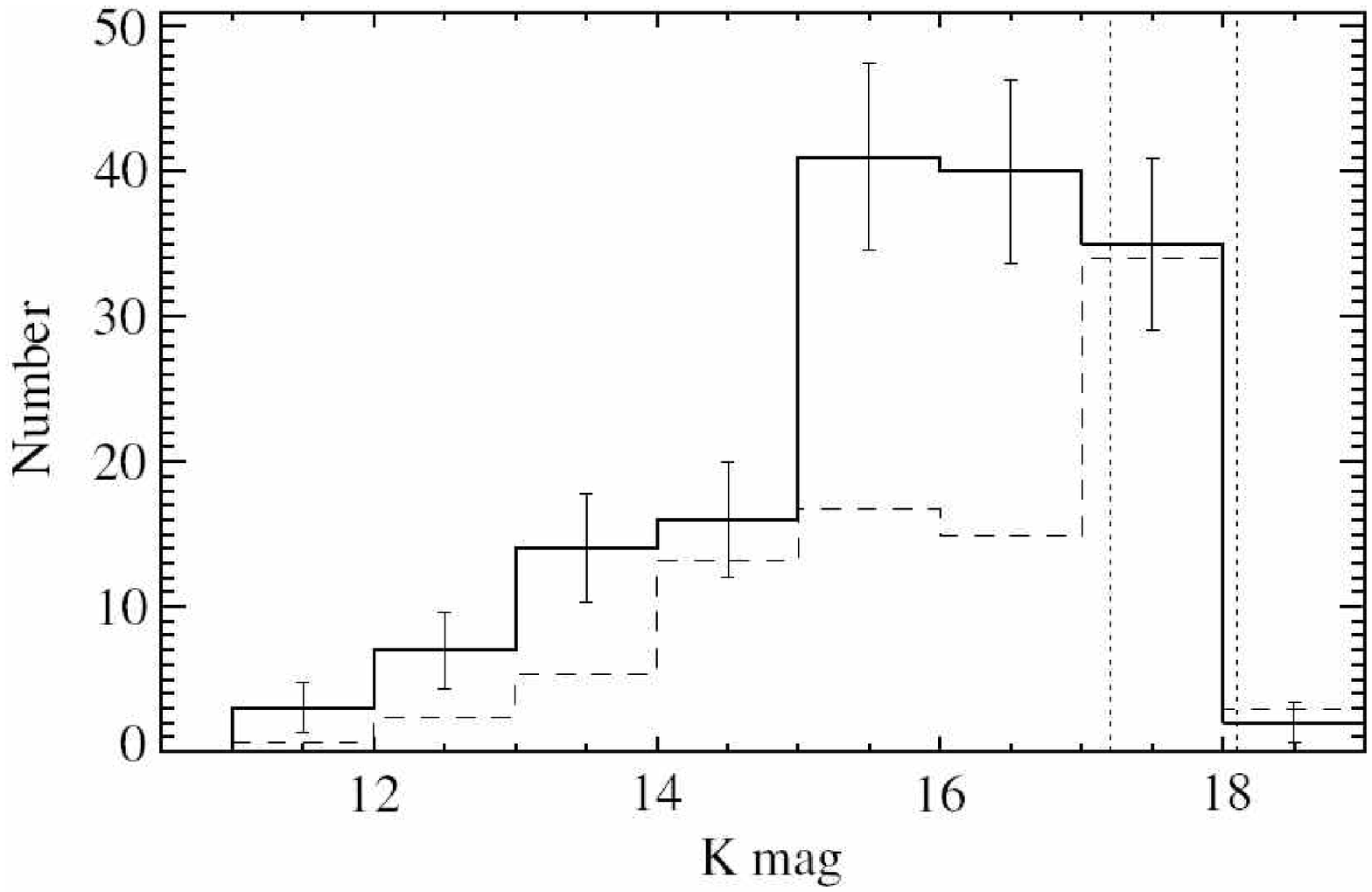}
\caption{The KLF of the Treasure Chest cluster, which includes only
stars redward of the 0.1 Myr isochrone in the CMD (Figure 8).  The
dashed histogram shows the corresponding KLF for stars in the
background annulus region, scaled to the same spatial area as the
cluster region.  Dotted vertical lines at $K$=17.2 and 18.1 represent
our completeness limit and limiting sensitivity, respectively.}
\end{figure}


\begin{thebibliography}{}

\bibitem[Bessell \& Brett(1988)]{bessbrett88}
Bessell, M.\ S., \& Brett, J.\ M.\ 1988, \pasp, 100, 1134

\bibitem[]{} Cardelli, J.A., Clayton, G.C., \& Mathis, J.S.\ 1989,
ApJ, 345, 245

\bibitem[D'Antona \& Mazzitelli(1997)]{dam97}
D'Antona, F.~\& Mazzitelli, I.\ 1997, Memorie della Societa 
Astronomica Italiana, 68, 807

\bibitem[]{} de Grauww, T., et al.\ 1981, A\&A, 102, 257

\bibitem[]{} Dutra, C.M., \& Bica, E.\ 2001, A\&A, 376, 434

\bibitem[]{} Ferland, G.J.\ 1996, Hazy, a brief introduction to {\sc
cloudy}, Univ.\ Kentucky Department of Physics and Astronomy Internal
Report

\bibitem[]{} Genzel, R., \& Stutzki, J.\ 1989, ARAA, 27, 41

\bibitem[]{} Ghosh, S.K., et al.\ 1988, ApJ, 330, 928

\bibitem[Haisch, Lada, \& Lada(2000)]{haisch00}
Haisch, K.\ E., Lada, E.\ A., \& Lada, C.\ J.\ 2000, \aj, 120, 1396

\bibitem[]{} Harvey, P.M., Hoffmann, W.F., \& Campbell, M.F.\ 1979,
ApJ, 227, 114

\bibitem[]{} Herbst, W.\ 1975, AJ, 80, 212

\bibitem[Hillenbrand \& Hartmann(1998)]{hill98}
Hillenbrand, L.\ A., \& Hartmann, L.\ W.\ 1998, \apj, 492, 540

\bibitem[]{} Hummer, D.G., \& Storey, P.J.\ 1987, MNRAS, 224, 801

\bibitem[Lada \& Lada(1995)]{lada95}
Lada, E.\ A., \& Lada, C.\ J.\ 1995, \aj, 109, 1682

\bibitem[]{} Massey, P., \& Johnson, J.\ 1993, AJ, 105, 980

\bibitem[]{} McCaughrean, M.J., \& Andersen, M.\ 2002, A\&A, 389, 513

\bibitem[]{} Megeath, S.T., Cox, P., Bronfman, L., \& Roelfsema, P.R.\
1996, A\&A, 305, 296

\bibitem[Meyer, Calvet, \& Hillenbrand(1997)]{meyer97} 
Meyer, M.~R., Calvet, N., \& Hillenbrand, L.~A.\ 1997, \aj, 114, 288

\bibitem[Muench et al.(2002)]{muench02}
Muench, A.\ A., Lada, E.\ A., Lada, C.\ J., \& Alves, J.\ 2002,
\apj, 573, 366

\bibitem[]{} Osterbrock, D.P.\ 1989, Astrophysics of Gaseous Nebulae
and Active Galactic Nuclei (Mill Valley: University Science Books)

\bibitem[]{} Rathborne, J.M., Brooks, K.J., Burton, M.G., Cohen, M.,
\& Bontemps, S.\ 2004, A\&A, 418, 563

\bibitem[]{} Reipurth, B., \& Bally, J.\ 2001, ARAA, 39, 403

\bibitem[]{} Salas, L., Rosado, M., Cruz-Gonzalez, I., Gutierrez, L.,
Valdez, J., Bernal, A., Luna, E., Ruiz, E., \& Lazo, F.\ 1999, ApJ,
511, 822

\bibitem[]{} Schild, H., Miller, S., \& Tennyson, J.\ 1997, A\&A, 318, 608

\bibitem[]{} Shuping, R.Y., Morris, M., \& Bally, J.\ 2004, AJ, 128, 363

\bibitem[]{} Smith, N.\ 2002, MNRAS, 331, 7

\bibitem[]{} Smith, N., \& Morse, J.A., 2004, ApJ, 605, 854

\bibitem[]{} Smith, N., Bally, J., \& Brooks, K.J.\ 2004a, AJ, 127, 2793 

\bibitem[]{} Smith, N., Bally, J., \& Morse, J.A.\ 2003, ApJ, 587, L105

\bibitem[]{} Smith, N., Egan, M.P., Carey, S., Price, S.D., Morse,
J.A., \& Price, P.A.\ 2000, ApJ, 532, L145

\bibitem[]{} Smith, N., Barb\'{a}, R.H., \& Walborn, N.R.\ 2004b,
MNRAS, 351, 1457

\bibitem[]{} Smith, R.G.\ 1987, MNRAS, 277, 943

\bibitem[]{} Sugitani, K., Tamura, M., Nakajima, Y., Nagashima, C.,
Nagayama, T., Nakaya, H., Pickles, A.J., Nagata, T., Sato, S., Kukuda,
N., \& Ogaura, K.\ 2002, ApJ, 565, L25

\bibitem[]{} Thackeray, A.D.\ 1950, MNRAS, 110, 524

\bibitem[]{} Thompson, R.I., Smith, B.A., \& Hester, J.J.\ 2002, ApJ,
570, 749

\bibitem[]{} van den Bergh, S., \& Herbst, W.\ 1975, AJ, 80, 208

\bibitem[]{} Walborn, N.R.\ 1995, RevMexAA, Ser.\ Conf., 2, 51

\bibitem[]{} Walborn, N.R.\ 2001, in Hot Star Workshop III: The
Earliest Phases of Massive Star Birth, ed.\ P.A.\ Crowther (San
Fracisco: ASP), 111

\bibitem[]{} Walborn, N.R., et al.\ 2002, AJ, 123, 2754

\bibitem[]{} Walsh, J.R.\ 1984, A\&A, 138, 380

\end{thebibliography}
\end{document}